\documentclass[prb,twocolumn]{revtex4}
\usepackage[dvips]{graphicx}
\usepackage{latexsym,amsmath,amssymb,bm,euscript,eufrak}

\def\etal{~\textit{et~al.}} % etal
\def\ra{\rangle} % bra
\def\la{\langle} % ket

\def\rt3rt3{{\sqrt{3} \times \sqrt{3}}}
\def\nacoo{\rm{Na$_x$CoO$_2$ }}

\def\fbar{\overline{f}}
\def\D{{\EuScript D}}

\def\bk{{\bm k}}
\def\be{{\bm e}}

\begin{document}

\title{Study of the triangular lattice $tV$ model near $x=1/3$}

\author{O. I. Motrunich and Patrick A. Lee}
\affiliation{Department of Physics, Massachusetts Institute of
Technology, Cambridge MA 02139}

\date{January 11, 2004}

\begin{abstract}
We study extended Hubbard model on a triangular lattice near
doping $x=1/3$, which may be relevant for the recently discovered
superconductor Na$_x$CoO$_2 \cdot y$H$_2$O.
By generalizing this model to $N$ fermionic species,
we formulate a meanfield description in the limit of large $N$.
In meanfield, we find two possible phases: a renormalized
Fermi liquid and a $\rt3rt3$ charge density wave state.
The transition between the two phases is driven by increasing
the nearest neighbor repulsion and is found to be first order
for doping $x=1/3$, but occurs close to the point of the local
instability of the uniform liquid.
We also study fluctuations about the uniform
meanfield state in a systematic $1/N$ expansion, focusing
on the residual interaction of quasiparticles and possible
superconducting instabilities due to this interaction.
Upon moving towards the CDW instability, the increasing charge
fluctuations favor a particular $f$-wave triplet state.
(This state was recently discussed by Tanaka\etal, cond-mat/0311266). 
We also report a direct Gutzwiller wavefunction study
of the spin-1/2 model.
\end{abstract}

\maketitle

%%%%%%%%%%%%%%%%%%%%%%%%%%%%%%%%%%%%%%%%%%%%%%%%%%%%%%%%%%%%
\section{Introduction}
Motivated by the recently discovered superconductivity\cite{Takada}
in Na$_x$CoO$_2 \cdot y$H$_2$O ($x \!\!\approx\!\! 1/3$) and the unusual
electronic properties\cite{Terasaki} of the \nacoo series,
we began a study of he $tV$ model
on a triangular lattice,\cite{chargefrustr}
focusing primarily on the renormalized Fermi liquid regime.
We pointed out that nearest neighbor repulsion can lead to
significant renormalization of the effective hopping amplitude,
which seems to be the case in these materials.
We also argued that strong repulsion drives this model into a
$\rt3rt3$ charge ordered state near commensurate
dopings $x=1/3$ and $2/3$.

We continue this study here and perform a systematic
slave boson meanfield analysis of the Fermi liquid close
to the charge density wave (CDW) state for dopings near $x=1/3$.
We also consider the residual interaction between the
quasiparticles and study which superconductivity channels are
favored in the $tV$ model on the triangular lattice.
Of particular interest here is possible enhancement of some
channels due to charge fluctuations upon approaching the
CDW phase.

Such studies of superconductivity due to residual interaction
in the models with strong local repulsion are familiar in the
high-$T_c$ field.  Scalapino\etal\cite{Scalapino} studied
the Hubbard model on a 3D cubic lattice in a random phase
approximation (RPA) and found that $d$-wave pairing becomes
attractive close to the spin density wave transition.
Kotliar and Liu\cite{Kotliar} studied the infinite-$U$ Hubbard model
on a 2D square lattice in a systematic large-$N$ treatment
and found that the residual interaction from the no-double-occupancy
constraint favors $d$-wave superconductivity close to half
filling.  More recently, McKenzie\etal\cite{McKenzie} applied the
analysis of Kotliar and Liu to extended Hubbard model on the
square lattice at quarter filling and found a transition to the
$\sqrt{2} \times \sqrt{2}$ CDW, which is driven by the nearest
neighbor repulsion.  Merino and McKenzie\cite{Merino}
studied superconducting instabilities of the Fermi liquid
near this transition.
% ??? and found that $d_{xy}$ state is favored.

It is of interest to perform similar studies on the triangular
lattice.  RPA treatment in the spirit of Scalapino\etal\cite{Scalapino}
has been done by Tanaka\etal\cite{Tanaka} very recently.  They find
that in the regime of interest for Na$_x$CoO$_2 \cdot y$H$_2$O,
a particular $f$-wave triplet channel is favored close to the
CDW instability.

Here, we report a Kotliar-Liu type analysis for the
triangular lattice, which has not been done so far.
The advantage of our approach over the RPA is that the treatment
of the strong on-site repulsion is better controlled.

\begin{figure}
\centerline{\includegraphics[width=3.0in]{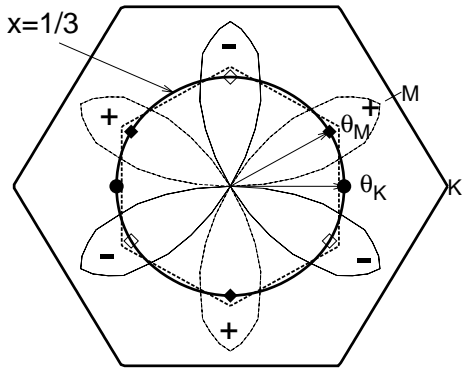}}
\vskip -2mm
\caption{Schematics of the nesting argument for doping $x=1/3$.
External hexagon shows triangular lattice Brillouin zone,
while internal hexagon shows reduced BZ for the
$\rt3rt3$ order.  Flat surfaces of the reduced BZ boundary
are connected by the $\rt3rt3$ ordering wavevectors
perpendicular to the surfaces.  Near these momenta transfer,
$V_{\rm eff}(\theta,\theta')$ becomes attractive close to the
CDW instability.
We also show the $f$-wave triplet gap function that we find
to be the dominant superconducting channel from the residual
interaction; this channel utilizes the near nesting of
$\theta_M$ points indicated with black diamonds.
}
\label{nesting}
\end{figure}

In our study, we also find that the preferred superconducting
channel near the transition to the $\rt3rt3$ CDW is
the $f$-wave triplet state discussed by Tanaka\etal\cite{Tanaka}.
This state has the lobes of large $\Delta$ oriented towards
$M$ points of the Brillouin zone (BZ) boundary as depicted
in Fig.~\ref{nesting}.
The origin of this result can be understood from the following
Fermi surface nesting argument.

In an RPA-type treatment, the effective interaction at
wavevector $q$ is
\begin{equation}
V_{\rm eff}(q) = \frac{V(q)}{1+ 2 \chi_0(q) V(q)}
\end{equation}
where $\chi_0(q)$ is positive.
This can be also written as
\begin{equation}
\label{veff}
V_{\rm eff}(q) = V(q) - 2 \chi(q) V(q)^2
\end{equation}
where $\chi(q) = \chi_0(q) / [1 + 2 \chi_0(q) V(q)]$ is the
full susceptibility in RPA.
Now, $V(q)$ is negative over some portion of the BZ,
in particular at the $\rt3rt3$ ordering wavevectors
such as $Q = 4\pi/(3a) \; \hat{\bm x}$,
so for sufficiently large $V$ there is an instability at which 
$\chi(Q)$ diverges.
Eq.~(\ref{veff}) shows that close to the instability the
effective interaction becomes strongly attractive at the
ordering wavevectors.
If there are sections of the Fermi surface that are nearly
connected by such wavevectors, then we expect enhanced
superconductivity in channels that utilize this nesting.
Our calculation using slave boson theory also produces attractive 
effective interactions due to charge fluctuations.  
The difference is that now the effective potential depends on 
${\bm k}$ and ${\bm k}^\prime$ of the electrons, rather than 
${\bm q} = {\bm k} - {\bm k}^\prime$.  
Near the Fermi surface we write this as 
$V_{\rm eff}(\theta, \theta^\prime)$ and we find that 
$V_{\rm eff}$ has attractive components when $\theta$ and 
$\theta^\prime$ are connected by an ordering vector $Q$.

Figure~\ref{nesting} shows the situation for doping $x=1/3$
(drawn to scale).
It turns out that at this filling the Fermi surface lies fairly
close to the reduced BZ corresponding to the $\rt3rt3$ order.
We also show the gap function $\Delta_k$ of the dominant
$f$-wave state.
This has large absolute values in the directions of the $M$
points of the full BZ boundary and signs as shown --- positive sign
near black diamonds and negative sign for white diamonds.  
With an attractive $V_{\rm eff}(Q)$, pairing between the 
black diamonds and also pairing between the white diamonds are 
very favorable, since the black diamonds are nearly connected
among themselves by ordering wavevectors, and so are the white diamonds.

We should of course consider the near nesting of other Fermi
surface segments as well.  For example, take the point $\theta_K$ 
and the point opposite to it on the Fermi surface shown by the filled 
circles in Fig.~\ref{nesting}.  These points are also nearly 
connected by an ordering vector and this pairing would favor a 
spin singlet state.
As an example, an $s$-wave state gains from all points $\theta,\theta'$
that are connected by $V_{\rm eff}(\theta,\theta')<0$.
The reason why the $s$-wave channel is still disfavored is that
there remains significant overall repulsion over non-nested
$V_{\rm eff}(\theta,\theta')$ that it cannot avoid.
The above $f$-wave state appears to be best for the overall 
$V_{\rm eff}$.

\vskip 2mm
The paper is organized as follows.  We first perform a slave
boson meanfield treatment of a generalized $tV$ model with $N$
fermion species.  Near $x=1/3$, two candidate states are
studied---a uniform Fermi liquid and a state with $\rt3rt3$
charge order.  The phase diagram is established.
We then study fluctuations over the uniform
saddle point in $1/N$ expansion, and focus on the residual
quasiparticle interactions from such fluctuations.
This approach can be viewed as a version of RPA that treats the
on-site constraint systematically, and provides a more quantitative
justification of the above nesting argument.

\vskip 2mm
We conclude this Introduction with one remark.  Earlier
works\cite{Baskaran1, KumarShastry, WangLeeLee, Ogata, Honerkamp}
considered $tJ$ model on the triangular lattice and found that the
dominant superconductivity from the $J$ interaction is $d+id$
singlet state.  This conclusion was also reached in our earlier 
report\cite{chargefrustr} in which we studied $tJV$ model and 
treated $J$ as the main residual interaction.
The $tV$ model with no $J$ terms predicts $f$-wave triplet state near 
the CDW order, and offers a way to distinguish between these two 
different pairing mechanisms.  The experimental situation regarding 
the pairing symmetry of Na$_x$CoO$_2 \cdot y$H$_2$O remains 
inconclusive, with controversy still surrounding whether there is a 
jump in the Knight shift below $T_c$.\cite{Waki,Kobayashi}

\section{Extended Hubbard model on triangular lattice}
\label{sec:meanfield}
We consider extended Hubbard model in the limit of large onsite
repulsion, i.e., the following $tV$ Hamiltonian
\begin{equation}
\hat H_{tV} = - P_G \sum_{ij} t_{ij} c_{i\sigma}^\dagger c_{j\sigma} P_G
+ \frac{1}{2} \sum_{ij} V_{ij} n_i n_j ~
\label{HtV}
\end{equation}
with nearest-neighbor repulsion $V_{ij} = V$.
$P_G$ projects out double occupation of sites.
The band is less than half-filled, with the average
fermion density of $1-x$ per site, and we specifically
consider the case $t>0$.
See Refs.~\onlinecite{Baskaran2, chargefrustr} for a more detailed
discussion of the possible application of this Hamiltonian to the
\nacoo system.
Here, we are primarily interested in the doping near $x=1/3$.

\subsection{Meanfield formalism}
Our meanfield treatment follows closely McKenzie\etal\cite{McKenzie}.
The general formalism is the same except for an arithmetic
difference in Eq.~\ref{Grr} and some minor differences in the analysis.
We apply this formalism to the triangular lattice $tV$ model.
Slave boson formulation is used to treat the no-double-occupancy
constraint.  We write $c_{i\sigma}^\dagger = f_{i\sigma}^\dagger b_i$,
and the slave boson Hamiltonian acts in the Hilbert space with
$f_{i\sigma}^\dagger f_{i\sigma} + b_i^\dagger b_i = 1$;
the boson field $b$ keeps track of the empty sites.

To formulate meanfield description and also in order to
go beyond the meanfield in a systematic manner, we consider
a generalized model with $N$ fermionic species.
The slave boson Hamiltonian is written as
\begin{equation}
\label{Hfb}
\hat H = -\frac{1}{NS} \sum_{ij} t_{ij}
  f_{i\sigma}^\dagger f_{j\sigma} b_j^\dagger b_i
+ \frac{1}{2NS} \sum_{ij} V_{ij} f_{i\sigma}^\dagger f_{i\sigma}
  (NS - b_j^\dagger b_j) ~,
\end{equation}
which now acts in the space
\begin{equation}
f_{i\sigma}^\dagger f_{i\sigma} + b_i^\dagger b_i = NS ~
\end{equation}
(the spin index $\sigma$ runs from $1$ to $N$).
For convenience, we introduced parameter $S$, which is kept
fixed as we take $N \to \infty$.  We also used a particular
form for the repulsion term.
To study the system behavior with doping, we fix the total
fermion number to $NS(1-x)$ fermions per site
(our large $N$ limit is thus like a thermodynamic limit in the
fermion flavors).
At the end of the calculation, we will put $N=2$, $S=1/2$,
and our specific choices when defining $H_{fb}$ are such
that this will reproduce the slave boson Hamiltonian for the
spin-1/2 model.

Proceeding as in McKenzie\etal\cite{McKenzie} and
Kotliar and Liu\cite{Kotliar}, we write the path integral in the
radial gauge\cite{ReadNewns}
\begin{equation}
Z = \int \D\fbar \D f \D r \D\lambda \;
\exp\left[-\int_0^\beta d\tau {\EuScript L}(\tau) \right] ~.
\end{equation}
The imaginary time Lagrangian
\begin{widetext}
\begin{equation}
\label{Lrad}
{\EuScript L} = \sum_i
\fbar_{i\sigma} (\partial_\tau - \mu + i\lambda_i) f_{i\sigma}
- \frac{1}{NS} \sum_{ij} t_{ij} r_i r_j \fbar_{i\sigma} f_{j\sigma}
+ \frac{1}{2NS} \sum_{ij} \fbar_{i\sigma}f_{i\sigma} V_{ij} (NS - r_j^2)
+ \sum_i  i\lambda_i (r_i^2 - NS) + \mu \sum_i NS(1-x) ~.
\end{equation}
\end{widetext}
To obtain this, auxiliary field $\lambda_i(\tau)$ was introduced
to enforce the constraint; the phase of the original boson field
$b_i$ was gauged away, while the absolute value is now
a real field $r_i = |b_i|$ [the corresponding measure is
$\D r = \prod_{i,\tau} r_i(\tau) dr_i(\tau)$].
Chemical potential $\mu$ sets the correct fermion density.

We now formally integrate out the fermions and seek saddle
points of the resulting action in terms of the fields $r_i$ and
$\lambda_i$.  The natural scale for the $r$ fields is $r^2 \sim N$,
and upon such rescaling (which we do not perform explicitly)
the saddle point analysis can be cast into a formal large $N$ procedure.

The saddle point conditions read
\begin{equation}
\la \fbar_{i\sigma} f_{i\sigma} \ra + r_i^2 = NS  ~,
\end{equation}
i.e., the constraints are satisfied on average, and
\begin{eqnarray}
\varphi_i \equiv i \lambda_i &=&
\frac{1}{2NS} \sum_j \frac{r_j}{r_i}
  \left[t_{ij} \la \fbar_{i\sigma} f_{j\sigma} \ra +
        t_{ji} \la \fbar_{j\sigma} f_{i\sigma} \ra \right]
\nonumber \\
&+& \frac{1}{2NS} \sum_j V_{ji} \la \fbar_{j\sigma} f_{j\sigma} \ra ~.
\end{eqnarray}
We seek time-independent saddle points.  However, the fields may
be spatially varying to allow for possible charge inhomogeneity.
In this case, $\varphi_i + \frac1{2NS} \sum_j V_{ij} (NS-r_j^2)$
is an effective potential on site $i$ (cf.~Eq.~\ref{Lrad}),
while $\frac1{NS} r_i r_j t_{ij}$ is an effective hopping amplitude in
such meanfield.
Among all saddle points, we are to take the one
that minimizes the free energy specified by Eq.~(\ref{Lrad}).

We first consider the uniform saddle point, which has
$r_i^2 \equiv b^2 = NS x$ and
\begin{equation}
\label{lambda}
i\lambda_i \equiv \varphi = \frac{1}{NS} \frac{N}{L^d} \sum_k
(\tilde{t}_k + \frac12 \tilde{V}_0) f(\xi_k) ~.
\end{equation}
Here,
$\tilde{t}_k = \sum_{r'} t_{rr'} e^{-ik(r-r')}$
% $= 2 t (\cos{\bm k}\cdot{\bm e}_1 + \cos{\bm k}\cdot{\bm e}_2
%         + \cos{\bm k}\cdot{\bm e}_3 )$,
and similarly for $\tilde{V}_k$ (we will often drop tildes when
the meaning is unambiguous);
% ${\bm e}_1$, ${\bm e}_2$, and ${\bm e}_3$ are unit triangular
% lattice vectors;
$L^d$ is the number of lattice sites;
$f(\xi) = 1/(e^{\beta\xi}+1)$ is the Fermi distribution;
and $\xi_k$ is the quasiparticle energy measured relative
to the Fermi level
\begin{equation}
\xi_k = -x \tilde{t}_k + \varphi + \frac12 \tilde{V}_0 (1-x) - \mu ~.
\end{equation}
The chemical potential is tuned so that
\begin{equation}
\frac1{L^d} \sum_k f(\xi_k) = S(1-x) ~.
\end{equation}

The uniform saddle point represents a renormalized Fermi liquid
with effective hopping $x t_{ij}$.  We recognize such hopping energy
renormalization as coming from the configurational constraints
imposed by the no-double-occupancy condition.  At this level,
the repulsive interaction leads only to the shift
in the bottom of the band.

\subsection{Fluctuations over the uniform state}
Away from half-filling and for small $V$, the uniform saddle point
has the lowest free energy.  We can establish the region of
its local stability by considering small fluctuations above the
uniform state.
Proceeding as in Refs.~\onlinecite{McKenzie, Kotliar}, we obtain
the following quadratic action for the fluctuations
$i \lambda_i = \varphi + i\delta\lambda_i$, $r_i = b (1+\delta r_i)$
\begin{widetext}
\begin{equation}
\label{S2}
S^{(2)} = \frac12
\sum_{q, \omega_n}
\left(\begin{matrix}
      \delta r(-q, -\omega_n) & \delta\lambda(-q, -\omega_n)
\end{matrix}\right)
\left(\begin{matrix}
      \Gamma_{rr} & \Gamma_{r \lambda} \\
      \Gamma_{\lambda r} & \Gamma_{\lambda\lambda}
\end{matrix}\right)
\left(\begin{matrix} \delta r(q, \omega_n) \\
                     \delta\lambda(q, \omega_n)
\end{matrix}\right) ~,
\end{equation}
where $\omega_n$ is bosonic Matsubara frequency, and the
``inverse RPA propagator'' is given by
\begin{eqnarray}
\Gamma_{rr}(q, \omega_n) &=& 2b^2 \varphi
- \frac{2b^2}{NS} \frac{N}{L^d} \sum_k
    (t_{k+q} + \frac12 \tilde{V}_0) f(\xi_k)
- \frac{b^4}{(NS)^2} \frac{N}{L^d} \sum_k
    \frac{f(\xi_{k+q}) - f(\xi_k)}{i\omega_n + \xi_k - \xi_{k+q}}
    (t_k + t_{k+q} + V_q)^2 ~, \label{Grr} \\
\Gamma_{r\lambda}(q, \omega_n) &=& \Gamma_{\lambda r}(q, \omega_n)
= i \left[2 b^2 + \frac{b^2}{NS} \frac{N}{L^d} \sum_k
            \frac{f(\xi_{k+q}) - f(\xi_k)}{i\omega_n + \xi_k - \xi_{k+q}}
            (t_k + t_{k+q} + V_q) \right] ~, \\
\Gamma_{\lambda\lambda}(q, \omega_n) &=& \frac{N}{L^d} \sum_k
    \frac{f(\xi_{k+q}) - f(\xi_k)} {i\omega_n + \xi_k - \xi_{k+q}} ~.
\end{eqnarray}
\end{widetext}
The above expressions coincide with Eq.~(17) in 
Ref.~\onlinecite{McKenzie} upon replacements 
$t/(NS) \to t/N$, $V/(NS) \to 2V/N$, except that the 
$\tilde{V}_0$ term in $\Gamma_{rr}$ replaces $\tilde{V}_k$ there.

We can write more compactly
\begin{eqnarray}
\Gamma_{rr} &=& \frac{2b^2}{NS} Y - \frac{b^4}{(NS)^2}
(V_q^2 X_0 + 2V_q X_1 + X_2) ~,\\
\Gamma_{r\lambda} &=& \Gamma_{\lambda r} =
i \left[2b^2 + \frac{b^2}{NS}(V_q X_0 + X_1) \right] ~,\\
\Gamma_{\lambda\lambda} &=& X_0 ~,
\end{eqnarray}
where we used Eq.~(\ref{lambda}) and introduced
\begin{eqnarray}
X_p(q, \omega_n) &=& \frac{N}{L^d} \sum_k (t_k + t_{k+q})^p
   \frac{f(\xi_{k+q}) - f(\xi_k)}{i\omega_n + \xi_k - \xi_{k+q}} ~,\\
Y(q) &=& \frac{N}{L^d} \sum_k (t_k - t_{k+q}) f(\xi_k) ~.
\end{eqnarray}

As discussed in Ref.~\onlinecite{Kotliar}, the local stability of the
saddle point is determined by the condition
$\det \Gamma = \Gamma_{rr} \Gamma_{\lambda\lambda}
- \Gamma_{r\lambda} \Gamma_{\lambda r} > 0$, which translates to
\begin{equation}
X_0 \left(\frac{4 b^2}{NS} V_q + \frac{2Y}{NS}
          - \frac{b^2}{NS} \frac{X_2}{NS} \right)
+ b^2 \left(2+\frac{X_1}{NS}\right)^2 > 0 ~.
\end{equation}

We now focus on the stability to static perturbations, setting
$\omega_n = 0$.
Since $X_0>0$ and $V_q$ is negative in some portion of the
Brillouin zone, there is clearly an instability for sufficiently
large $V$.

\begin{figure}
\centerline{\includegraphics[width=\columnwidth]{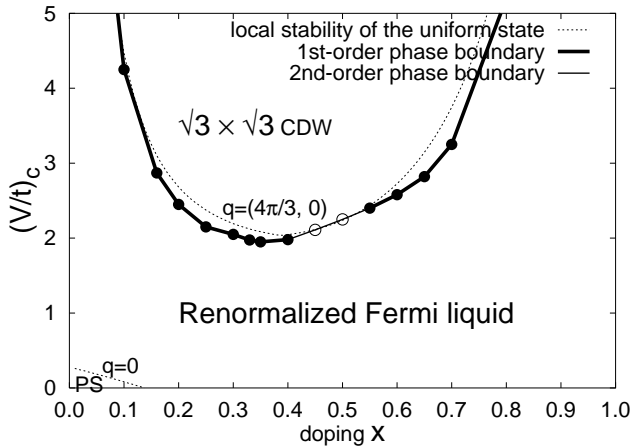}}
\vskip -2mm
\caption{Meanfield phase diagram of the $tV$ Hamiltonian.
For $x<0.14$, the uniform state is unstable at $q=0$ 
(towards phase separation) for small repulsion $V$, 
but is stabilized for moderate $V$.
For large $V$, the instability occurs at the $\rt3rt3$ ordering
wavevector and the critical $(V/t)_c$ is shown with the dotted
line.  In the region between the dotted lines, the uniform
state is locally stable at all wavevectors.
The circles show the actual meanfield transitions with general
$\rt3rt3$ ansatz; filled circles indicate first-order
phase transitions, while open circles indicate second-order transitions.
}
\label{mfphased}
\end{figure}

The local stability analysis specialized to $N=2, S=1/2$
(and in the zero-temperature limit) for all dopings is
summarized in Fig.~\ref{mfphased}.

For $x<0.14$ there is an instability at $q=0$ for small $V$,
which persists as long as
\begin{equation}
\label{PS}
\det\Gamma(q=0) = {\rm const} \times
[ 1 + 2 t_F \nu(\epsilon_F) + \tilde{V}_0 \nu(\epsilon_F) ] < 0 ~,
\end{equation}
where we used $X_0(q\!\!=\!\!0) = \nu(\epsilon_F)$, etc.,
$\nu(\epsilon_F)$ is the spinful density of states per site
at the Fermi level.
The instability is towards phase separation into hole-rich
and hole-poor regions.  This can be seen by examining the energy
of the uniform state (measured per site)
\begin{equation}
H(x) = E_t(x) + \frac{1}{2}\tilde{V}_0 (1-x)^2 ~,
\end{equation}
where $E_t(x)$ is the hopping energy per site in the free fermion
problem with hopping amplitude $tx$ and fermion density $(1-x)$.
After some analysis, the condition~(\ref{PS}) is seen to be
equivalent to $H''(x)<0$, which indeed leads to phase separation.
Note that moderate nearest-neighbor repulsion $V$ stabilizes the uniform
saddle point against phase separation, and the corresponding
boundary is shown in the lower left hand corner of Fig.~\ref{mfphased}.

The instability of the uniform state for large $V$ is found
to always occur at the $\rt3rt3$ ordering wavevector.
The corresponding critical $(V/t)_c$ is shown with the dotted line
in Fig~\ref{mfphased}.
In the region designated renormalized Fermi liquid the uniform
state is stable towards static fluctuations at any wavevector.

\vskip 2mm

\subsection{$\rt3rt3$ CDW saddle points}
However, local stability does not guarantee global stability of the
saddle point.  As a specific example, we consider $x=1/3$.
The above local stability analysis gives
\begin{equation}
V_c (x=1/3) = 2.11 ~.
\end{equation}

The uniform saddle point has the energy per site
\begin{equation}
E_{\rm uniform}(x=1/3) = - 6 t x \la f_{i\sigma}^\dagger f_{j\sigma} \ra
+ 3 V (1-x)^2
\end{equation}
with $\la f_{i\sigma}^\dagger f_{j\sigma} \ra = 0.337$ for the
uniform hopping at this doping.

On the other hand, consider a competing state with complete
$\rt3rt3$ order which has all charges on the $B$ and $C$
sublattices, $b_B=b_C=0$, while the sublattice $A$ is not occupied,
$b_A=1$.  The energy per site is
\begin{equation}
E_{\rt3rt3}(x=1/3) = V.
\end{equation}
It is simple to see that the latter becomes energetically
preferred over the uniform state for $V=2.00$, i.e.,
below the local instability point of the uniform state,
which means that the transition is first order.
The following detailed analysis finds that the
the actual transition occurs at $V_c^{\rm 1st order}(x=1/3)=1.98$
to a state which is close to the above state with
complete CDW order.

To study such possibilities in detail, we consider saddle points
that have the $\sqrt{3} \times \sqrt{3}$ ordering pattern.
The procedure is as follows:
We select the $A$-sublattice of the three sublattices,
and seek saddle points that have $b_i = b_A$ on the $A$ sublattice
and $b_i=b_B=b_C$ on the $B$ and $C$ sublattices.
Of course, these must satisfy $b_A^2 + b_B^2 + b_C^2 = 3x$.
Similarly, we allow different $\varphi_A$ and $\varphi_B=\varphi_C$.
For convenience, we can take
$\varphi_A + \varphi_B + \varphi_C = 0$,
since there is also the chemical potential degree of freedom
which is tuned to obtain the correct fermion density for
each set $b_A, \varphi_A$.  Now, for each $b_A$ we tune $\varphi_A$
until $b_A^2 + \la f_{A\sigma}^\dagger f_{A\sigma} \ra = 1$
is satisfied, and finally adjust $b_A$ so that the other
self-consistency condition is satisfied.
We find all such saddle points and compare their free energies
obtained from Eq.~(\ref{Lrad}).
The result is shown in Fig.~\ref{mfphased} with circles.
Filled circles indicate the situation similar to $x=1/3$, when
the transition is first order.  For open circles, we tentatively
conclude that the transition point coincides with that determined
by the local stability analysis and the transition is second order.

\vskip 2mm
We considered in detail the uniform and the $\rt3rt3$ CDW
saddle points.  Since we are interested in the doping regime
near $x=1/3$, where we do not expect some other state to enter
the competition, the presented meanfield analysis is complete.

%%%%%%%%%%%%%%%%%%%%%%%%%%%%%%%%%%%%%%%%%%%%%%%%%%%%%%%%%%%%%%%%%%
\section{Residual Superconducting instabilities}
We now aim to go beyond the meanfield description.
The free energy of the uniform state at next order in $1/N$ can
be obtained from the quadratic action $S^{(2)}$, Eq.~(\ref{S2}).
In terms of the fermionic quasiparticles, this contains further
effective mass renormalization as well as residual interaction of
quasiparticles mediated by the bosons.
We focus on the residual interaction, in order to decide
which pairing channel is favored at low temperatures for our
triangular lattice with hard repulsion.

The vertices coupling the fermions and the bosons can be
obtained by examining the Lagrangian Eq.~(\ref{Lrad}),
while the boson propagators are given by the inverse of the
matrix $\hat\Gamma(q, \omega_n)$ in the quadratic action,
e.g., $D_{rr}(q, \omega_n) \equiv
\la \delta r(-q, -\omega_n) \delta r(q, \omega_n) \ra
= \Gamma_{\lambda\lambda}/\det\hat\Gamma$, etc.
From the expressions for $\Gamma$ and using $b^2 \sim N$,
all propagators are $O(1/N)$.

For our study, we single out interaction terms relevant for the
BCS instability, obtaining
\begin{widetext}
\begin{eqnarray}
\label{Veff12}
V_{\rm eff}(k_1, k_2) = D_{\lambda\lambda}
 - \frac{b^4}{(NS)^2} D_{rr} |t_{k_1} + t_{k_2} + V_q|^2
 + i \frac{b^2}{NS} \left[D_{\lambda r} (V_{-q} + t_{-k_1} + t_{-k_2})
                      + D_{r\lambda} (V_q + t_{k_1} + t_{k_2}) \right] ~,
\end{eqnarray}
\end{widetext}
where all boson propagators are at wavevector $q=k_1-k_2$.
The interaction is of order $O(1/N)$.
If we ignore the frequency dependence of the boson propagators,
the relevant terms correspond to the standard pairing Hamiltonian
\begin{equation}
\label{Hpairing}
\hat{H}_{\rm pairing} = \frac{1}{2 L^d} \sum_{k_1, k_2}
V_{\rm eff}(k_1, k_2) f_{k_1,\sigma}^\dagger f_{-k_1,\sigma'}^\dagger
                  f_{-k_2,\sigma'} f_{k_2,\sigma} ~.
\end{equation}

\begin{figure}
\centerline{\includegraphics[width=\columnwidth]{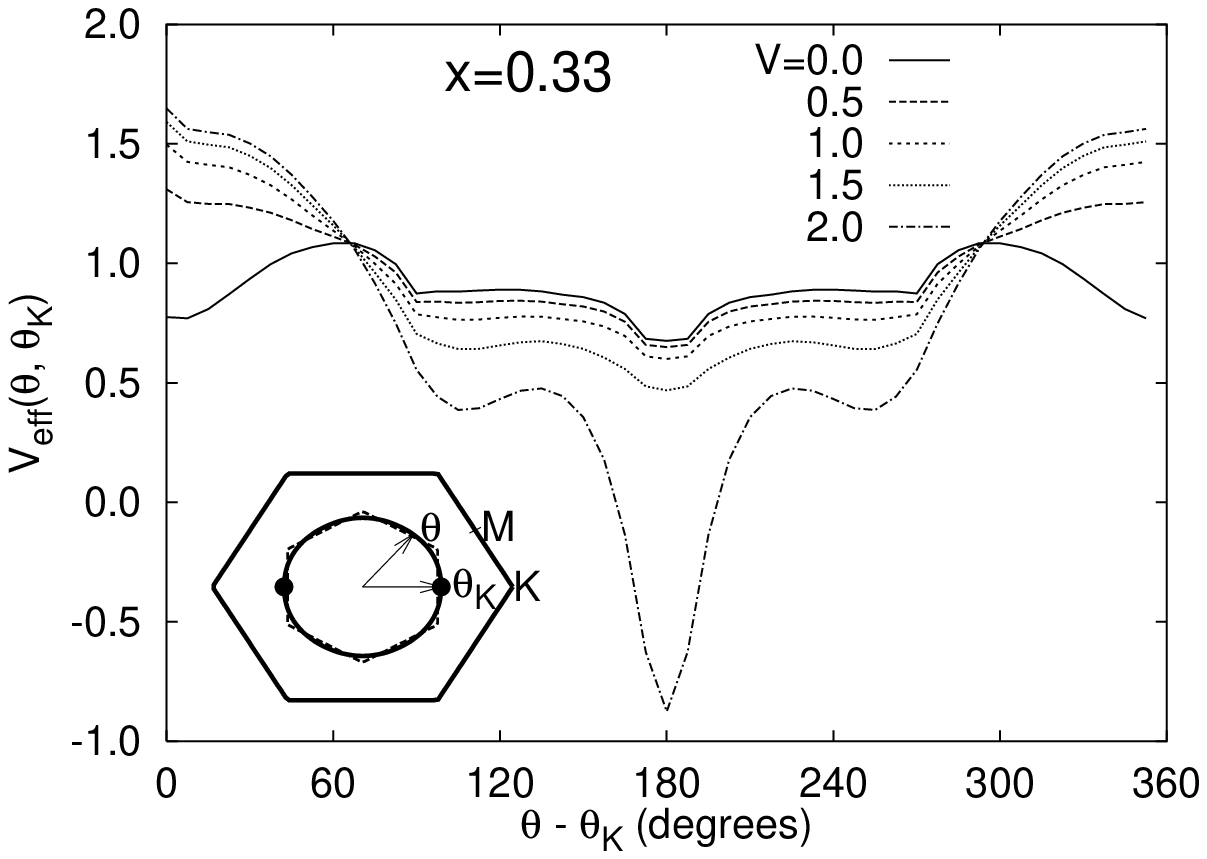}}
\centerline{\includegraphics[width=\columnwidth]{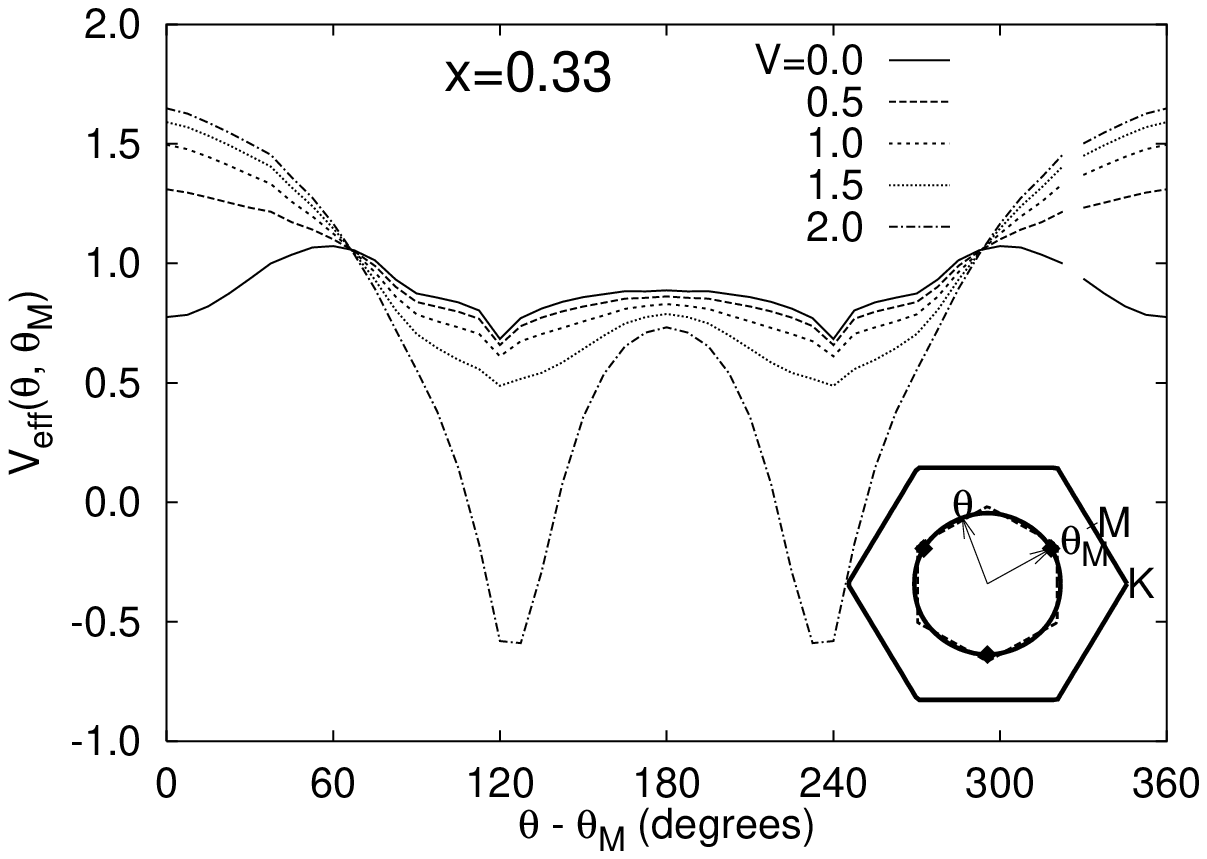}}
\vskip -2mm
\caption{Residual quasiparticle interaction relevant
for the pairing instability (cf.~Eq.~\ref{Hpairing})
plotted for the momenta $k_1$ and $k_2$ on the Fermi surface.
One angle is kept fixed while the other is varied over the FS
as indicated in the insets, which also show the triangular
lattice BZ and the reduced BZ for the $\rt3rt3$ CDW order
(see~Fig.~\ref{nesting}).  The data is for $x=1/3$, and the
uniform state becomes unstable at $V_c=2.11$.
The vertical scale is in units of $t$.
Top panel:  The fixed angle is in the direction of the $K$
point of the full BZ boundary.  As we approach the CDW instability,
$180^\circ$ scattering amplitude becomes attractive, which
is associated with the fact that the corresponding
Fermi surface points (black circles) are nearly connected
by one of the $\rt3rt3$ ordering wavevectors.
Bottom panel:  The fixed angle is in the direction of the $M$
point; the three nearly nested points occur at $120^\circ$
from each other.
}
\label{fig:veff}
\end{figure}

From now on, we fix $N=2$, $S=1/2$.
We first consider in detail doping $x=1/3$.
%(note that this doping is not special in our formalism).
We are primarily interested in the scattering processes when
momenta $k_1$ and $k_2$ lie near the Fermi surface.
Therefore, we visualize the effective interaction by fixing the
momenta on the Fermi surface and plotting
$V_{\rm eff}(\theta_1, \theta_2)$ as a function
of polar angles.  This is shown in Fig.~\ref{fig:veff},
where we see how $V_{\rm eff}$ evolves for increasing $V$
approaching the CDW transition.
Observe that there is residual repulsion even when $V=0$,
which is entirely due to the no-double-occupancy constraint,
and that this repulsion already has some momentum space features.
As we increase $V$ from zero, the initial effect is to
add more repulsion at zero momentum transfer.
Increasing $V$ further and approaching the $\rt3rt3$ CDW instability,
there is a dramatic development in the features associated with
scattering at the $\rt3rt3$ ordering wavevectors.
As discussed in the introduction, the quasiparticle interaction at such
momentum transfer becomes attractive close to the critical point,
and this is clearly seen in the plots.
In Fig.~\ref{fig:veff}, the critical point $V_c=2.11$ is
defined from the quadratic fluctuation analysis.  As discussed earlier,
in the meanfield, we find instead $1$st-order phase transition at a
somewhat lower $V_c^{\rm 1st order}=1.98$.  However the features in
$V_{\rm eff}(\theta, \theta')$ are already enhanced at the actual
transition since it occurs close to the instability point and
has significant critical fluctuations, so we will mostly ignore
the distinction in what follows.

We now study BCS instabilities due to this residual interaction;
the analysis below is valid for either singlet or $S_z=0$
triplet superconducting channels.
% The finite temperature gap equation reads
% \begin{equation}
% \Delta_k = -\frac{1}{L^d} \sum_{k'} V_{\rm eff}(k, k')
% \frac{\tanh(\beta E_{k'}/2)}{2E_{k'}} \Delta_{k'}
% \end{equation}
% where $E_k = \sqrt{\xi_k^2 + |\Delta_k|^2}$.
We follow Refs.~\onlinecite{Scalapino, Kotliar}
and define coupling constant associated with each channel
$\Delta_k \sim g_\alpha(k)$
\begin{equation}
\label{eigval}
c_{\alpha} = \,\, {\bm -} \,\,
              \frac{\displaystyle \int \frac{d\sigma_k}{|v_k|}
                    \int \frac{d\sigma_{k'}}{(2\pi)^d |v_{k'}|}
                    g_\alpha(k)^* V_{\rm eff}(k, k') g_\alpha(k')}
                   {\displaystyle \int \frac{d\sigma_k}{|v_k|}
                                       |g_\alpha(k)|^2} ~,
\end{equation}
where the integration is over the Fermi surface elements $d\sigma$,
and $v(k) = \nabla_k \xi(k)$ is the Fermi velocity.
For an attractive channel, we must have $c_\alpha > 0$, and the 
transitions temperature is given by the weak coupling BCS
expression (which is appropriate in the present case)
\begin{equation}
\label{Tc}
T_c[\alpha] = \Omega \exp[-1/c_\alpha] ~;
\end{equation}
the frequency cutoff is roughly $\Omega \sim x t$,
since the energy integration is over the entire band.

Representative triangular lattice tight binding harmonics that
cover main symmetry classes are listed in Table~\ref{tab:gs}.
Four of the ansatze have their real-space $\Delta_{rr'}$
nonzero on nearest-neighbor bonds only,
the NNN $f$-wave ansatz has next-nearest-neighbor bonds,
and the $i$-wave ansatz requires even further neighbor bonds.
Since $V_{\rm eff}$ is real, only real ansatze need to be considered.
Indeed, in this case the problem of finding a harmonic with the
largest coupling constant is equivalent to a real symmetric
eigenvalue problem; the easiest way to see this is to discretize
the $d\sigma$ integrals and define
$\psi_\alpha(k) = g_\alpha(k) (d\sigma_k/|v_k|)^{1/2}$.

\begin{table}
\begin{tabular}{|c|c|}
\hline\hline
Ansatz $g_\bk$  & Label \\
\hline\hline
$\cos\bk\cdot\be_1 + \cos\bk\cdot\be_2 + \cos\bk\cdot\be_3$ & $s$ \\
\hline
$\sin\bk\cdot\be_2 - \sin\bk\cdot\be_3$ & $p_x$ \\
\hline
$\cos\bk\cdot\be_2 - \cos\bk\cdot\be_3$ & $d_{xy}$ \\
\hline
$\sin\bk\cdot\be_1 - \sin\bk\cdot\be_2 + \sin\bk\cdot\be_3$ & $f$ \\
\hline
$\sin\bk\cdot(\be_1+\be_2) - \sin\bk\cdot(\be_2+\be_3)
+ \sin\bk\cdot(\be_3-\be_1)$ & NNN $f$ \\
\hline
$\cos\bk\cdot(2\be_1 \!+\! \be_2) - \cos\bk\cdot(\be_1 \!+\! 2\be_2) +
 \cos\bk\cdot(2\be_2 \!+\! \be_3)$ & $i$ \\
$- \cos\bk\cdot(\be_2 \!+\! 2\be_3) + \cos\bk\cdot(2\be_3 \!-\! \be_1) -
 \cos\bk\cdot(\be_3 \!-\! 2\be_1) $ & \\
\hline
\end{tabular}
\caption{Triangular lattice harmonics evaluated in Fig.~\ref{fig:lambda}.
$\be_1$, $\be_2$, $\be_3=\be_2-\be_1$ refer to unit triangular
lattice vectors.  The labels are to be taken as descriptive only.
}
\label{tab:gs}
\end{table}

\begin{figure}
\centerline{\includegraphics[width=\columnwidth]{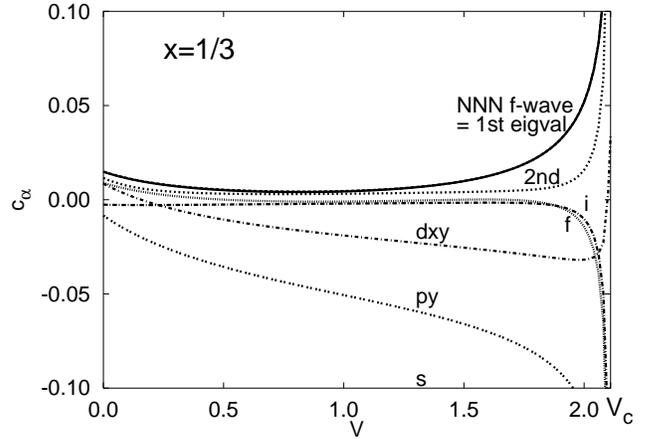}}
\vskip -2mm
\caption{Coupling constants for the triangular lattice
harmonics in Table~\ref{tab:gs}
(the $s$-wave ansatz is below the bottom of the plot).
The doping is $x=1/3$, and the critical $V_c=2.11$ sets
the right-hand plot boundary.
We also show the maximal and the second eigenvalue;
the NNN $f$-wave coupling constant coincides with the
maximal eigenvalue for all $V$ shown.
}
\label{fig:lambda}
\end{figure}

Figure~\ref{fig:lambda} shows the coupling constant $c_{\alpha}$
evaluated for the harmonics in Table~\ref{tab:gs}.
%(When we use $V_{\rm eff}$ from Eq.~\ref{Veff12} in the expression
%for $c_\alpha$, we also need to take into account the unit cell 
%volume factor when going from lattice sums to integrals.)
We find that for all $V$ (except very close to $V_c$)
the NNN $f$-wave triplet is the dominant channel,
while the other ansatze are repulsive or become
repulsive upon increasing $V$.
The performance of the NNN $f$-wave is understood by
looking at the bottom panel of Fig.~\ref{veff}.
The three marked points $120^\circ$ from each other are
nearly connected by $\rt3rt3$ ordering wavevectors and
the scattering among these points becomes attractive upon
approaching the CDW regime.
As displayed in Fig.~\ref{nesting}, the NNN $f$-wave ansatz
has its positive lobes oriented precisely in these three
directions and is able to utilize this attraction.

Here, we note that generic next-nearest-neighbor ansatze utilize
this nesting better than the nearest-neighbor ones, and we tried
the corresponding NNN versions of the $s$, $p$, and $d$-wave;
the coupling constants are improved (not shown),
but are still far from approaching the NNN $f$-wave ansatz.
This is because each ansatz also has to pay the cost of
repulsive scattering on large non-nested portions of the
Fermi surface, and the NNN $f$-wave ansatz appears to be
best here as well.  For example, the coupling constant for the
$s$-wave remains negative and appears on the scale of
Fig.~\ref{fig:lambda} only very close to the critical point,
despite the fact that it gains from all
attractive $V_{\rm eff}(\theta,\theta')<0$.

In fact, we also solve the full eigenvalue problem specified by
Eq.~(\ref{eigval}), and plot the first (maximal) and the second
eigenvalue in the same figure.
We find that the NNN $f$-wave coupling constant coincides with the
maximal eigenvalue in the entire range of $V<V_c$ 
(except maybe very close to $V_c$).  
The second eigenvalue is well separated from the first; 
the corresponding eigenvector over a large range of $V$ 
does not have one of the simple $s$, $p$, $d$ or $i$ character, 
but instead has eight lobes in momentum space, four of each sign.

\begin{figure}
\centerline{\includegraphics[width=\columnwidth]{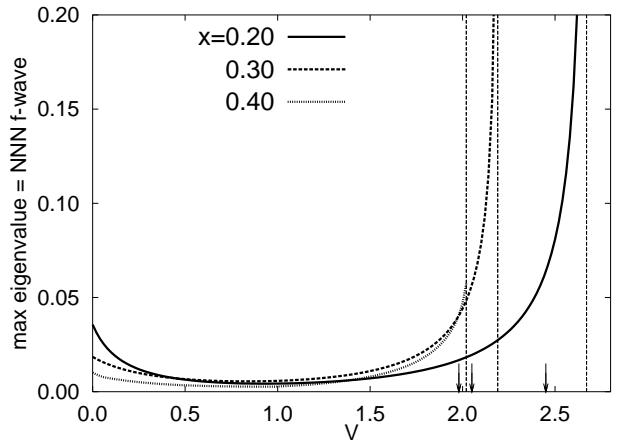}}
\vskip -2mm
\caption{Maximal eigenvalue for dopings $x=0.20$, $0.30$, and $0.40$.
The critical $V_c$ from the quadratic fluctuation analysis is
indicated by vertical line in each case, while the $1$st-order
transition is indicated with the corresponding arrow near the
bottom of the graph (cf.~Fig.~\ref{mfphased}).
}
\label{fig:lmbdallx}
\end{figure}

Finally, Fig.~\ref{fig:lmbdallx} shows the maximal eigenvalue
as a function of $V$ for several dopings.
In each case, the maximal eigenvalue corresponds
to the NNN $f$-wave ansatz.  The nesting displayed
in the bottom panel of Fig.~\ref{fig:veff} that favors the 
NNN $f$-wave improves as we lower the doping, since the Fermi 
surface passes through the end points of the reduced BZ 
when $x=0.194$ (see also Fig.~\ref{nesting}).
From Fig.~\ref{fig:lmbdallx} we see that the enhancement in the 
coupling constant is strongest and over the broadest range for $x=0.20$.
Observe also that the residual interaction from the
no-double-occupancy constraint only ($V=0$) also favors
the discussed NNN $f$-wave channel on the triangular lattice.

\vskip 2mm
We conclude by discussing what these results mean for the
scale of superconductivity.  We see that because of the
nesting the maximal $c_\alpha$ can be relatively large
compared to similar predictions in the non-nested cases
(such as square lattice at finite doping,
Ref.~\onlinecite{Kotliar}).
However, if we use Eq.~(\ref{Tc}) literally, the obtained scale of 
$T_c$ is still very small:
for example, if we take $c_\alpha=0.1$, which is fairly large,
then $\exp[-1/c_\alpha] \sim 5 \cdot 10^{-5}$, which is tiny.
Here we remark that one specific ingredient that can enhance $T_c$ 
is missing in the present treatment, namely, the enhancement of the 
effective mass by the nearest neighbor repulsion while 
remaining in the Fermi liquid.
More precisely, at the presented level of analysis,
the effective hopping $t_{\rm eff} = xt$ is not renormalized
by $V$, whereas it can become several times smaller when
additional short-range correlations are included
as discussed in Ref.~\onlinecite{chargefrustr}
and in the following section.
Since $1/c_\alpha \sim t_{\rm eff}/V_{\rm eff}$, this can have
dramatic effect on the calculated $T_c$.

\section{Jastrow-Gutzwiller wavefunction study.  Discussion}
We now consider in some detail how the above results apply to the
original spin-1/2 model.  To this end, we have performed a systematic
trial wavefunction study of the $tV$ model.  Specifically,
we consider a family of Jastrow-Gutzwiller
wavefunctions\cite{chargefrustr, Ceperley}
\begin{equation}
\Psi_{JG}(r_1\sigma_1, \dots) =
e^{-\sum_{i<j} u(r_i-r_j)}
\det\left[\psi_\alpha(R_j)\right]
\det\left[\psi_\alpha(R_j^\prime) \right] ~,
\end{equation}
where $\{R\}$ and $\{R'\}$ denote the positions of spin-up
and spin-down fermions respectively, and the wavefunction is
nonzero only when the two sets do not overlap, which is
the result of Gutzwiller projection.
$\{\psi_\alpha\}$ refer to appropriate single-particle states
that are occupied in the ``preprojected'' wavefunction.
Each configuration of fermions is also weighted by a Jastrow factor
defined via two-particle pseudopotentials $u(r_i-r_j)$,
which puts additional correlations into the wavefunction.

The Fermi liquid state is obtained by occupying appropriate plane
wave states $\psi_k$, $|k|\leq k_F$.
From our studies of the $tV$ model, we conclude that
already the nearest-neighbor Jastrow factor gives good control
over local correlations and is sufficient for an accurate
energetics distinction between the Fermi liquid and competing
charge ordered state.

As discussed at length in Ref.~\onlinecite{chargefrustr}, lattice gas
system described by the nearest-neighbor $u(r_i, r_j) \equiv W$
undergoes a transition to a $\rt3rt3$ state for large $W$.
Our direct optimization with such single-parameter
wavefunction showed that this state is driven into the
charge-ordered regime for sufficiently strong repulsion $V$
and dopings in the range $0.27 < x < 0.50$.
We also mentioned that once this happens, we are no longer
justified in using plane waves for the preprojected orbitals.
We should instead consider more general single-particle states;
for example, we consider orbitals obtained by diagonalizing a ``trial''
Hamiltonian
\begin{equation}
\hat{H}_{\rm CDW} = -\sum_{ij}
\chi_{ij} c_{i\sigma}^\dagger c_{j\sigma}
+ \sum_i \varphi_i c_{i\sigma}^\dagger c_{i\sigma} ~,
\end{equation}
where we allow generic hopping amplitudes $\chi_{ij}$
and site potentials $\varphi_i$ that follow the $\rt3rt3$
ordering pattern.
Specifically, we select the $A$ sublattice and take
$\chi_{AB}=\chi_{AC} = 1+\kappa$, $\chi_{BC} = 1-\kappa$;
$\kappa>0$ makes the hopping more dice-lattice-like, while
$\kappa<0$ makes it more honeycomb-like.  We also use the
convention $\varphi_B=\varphi_C = -\varphi_A/2$ and vary $\varphi_A$.
Observe that this structure of the trial Hamiltonian is
suggested by the meanfield treatment of the Fermi liquid - CDW
competition in Sec.~\ref{sec:meanfield}, and in fact by any such
meanfield.
However, our wavefunctions have further local correlations
built in by the Jastrow factors which allow the liquid to
better accommodate the nearest-neighbor repulsion
(also leading to significant renormalization of the bandwidth
as discussed in Ref.~\onlinecite{chargefrustr}),
while this is missing in the meanfield.

We summarize our wavefunction studies for three dopings
$x=0.24$, $0.33$, and $0.40$.  We report only the result
of the three-parameter $W,\kappa,\varphi_A$ optimization.
When $\kappa=0, \varphi_A=0$, we obtain the Fermi liquid state.
As discussed earlier, this restricted study should suffice
for an accurate determination of the uniform liquid to 
CDW transition.

For dopings $x=0.33$ and $0.40$, we find rather abrupt
transitions to the $\rt3rt3$ state at $V_c(x=0.33) \approx 3.5$
and $V_c(x=0.40) \approx 4.0$.
On the other hand, for the doping $x=0.24$, the system shows
no ordering till very large $V=15$ to $20$ and maybe even
higher.  Comparing with Fig.~\ref{mfphased}, the behavior
at this lowest doping is very different from the meanfield
prediction.  This is an extreme manifestation of the
fact that the charges are able to effectively avoid each other
and remain in the liquid state, which is captured by our
Jastrow-Gutzwiller wavefunction, while the meanfield uniform
state cannot accommodate this and pays large repulsion energy cost.

Finally, having established the dominant local energetics,
we also tried adding superconducting correlations on top
of the renormalized Fermi liquid state.  We tried several
superconducting ansatz including extended $s$-wave, $p+ip$,
$d+id$, $f$-wave, and also their NNN versions.
However, in our studies we were not able to detect any
improvement in the energetics upon adding superconductivity.
From this we conclude that if some such state appears
at low energies, the condensation energy is still very small
to be detected by direct numerical studies.
In this situation, we are left to rely on approximate
analytical calculations such as discussed in the main
body of this paper.

\acknowledgements
We thank Stuart Brown for bringing Ref.~\onlinecite{Merino} to our 
attention.  This work was supported by the National Science Foundation 
under grant DMR--0201069.
OIM also wants to thank his family for support during his stay in
Ukraine.

\end{document}